\documentclass{desyproc}

\begin{document}
\title{Hidden Higgs Scenarios: \\
new constraints and prospects at the LHC}


%

%

%
\author{{\slshape Radovan Derm\'\i \v sek}  \\[1ex]
Department of Physics, Indiana University, Bloomington, IN 47405, USA}

\contribID{xy}  
\confID{1964}
\desyproc{DESY-PROC-2010-01}
\acronym{PLHC2010}
\doi            

\maketitle

\begin{abstract}
We review the motivation for hidden Higgs scenarios and discuss the light CP odd Higgs scenario in the NMSSM as an example. We summarize experimental constraints  including recent limits from BaBar and Aleph. The main part of the talk is the discussion of dominant decay modes of the standard model like Higgs boson, and related decay modes of the charged Higgs and heavy CP even Higgs bosons, in these scenarios with the focus on signatures and prospects for the LHC. Examples include the direct production of a light CP odd Higgs boson, and a light charged Higgs boson in top quark decays.
\end{abstract}

\section{Motivation for non-standard Higgs  decays}

One of the most important questions in particle physics is: Where is the Higgs boson? The LEP exclusion limits, $m_h > 114$ GeV, constraints from precision electroweak data, $m_h < 157$ GeV~\cite{lepewwg}, and recently also the Tevatron limits,  leave about 40 GeV window for the standard model (SM) Higgs boson.
This 40 GeV window is very interesting and there are several suggestive hints or coincidences related to it. First of all, this window overlaps with the range of  Higgs masses in which the standard model  can be a consistent theory all the way to the grand unification scale or the Planck scale, $m_h \simeq 125 - 175$ GeV. Another interesting coincidence is that this window also overlaps with the range of Higgs masses predicted in the minimal supersymmetric model (MSSM), $m_h \lesssim 135$ GeV. 
 Hence, it is expected that the  Higgs boson is somewhere in this window and most of the effort is focused on discovery strategies related to this possibility. 
 
 However there are also several compelling hints that the SM-like Higgs boson is below the LEP exclusion  limits. First of all, electroweak symmetry breaking (EWSB) in  simple supersymmetric (SUSY) extensions of the standard model, with superpartners near the electroweak (EW) scale, generically predicts the Higgs boson  not heavier than about 100 GeV. Non-observation of the Higgs boson at LEP resulted in the ``fine-tuning" problem in these models~\cite{Dermisek:2009si}. Second of all, the best fit to  precision electroweak data is achieved for the Higgs mass of 87 GeV~\cite{lepewwg}. The third hint comes from the LEP data: the largest excess of Higgs like events at LEP corresponds to the Higgs mass of 98 GeV.
It is an interesting coincidence that natural EWSB in SUSY models, precision EW data, and the largest excess of Higgs like events point to the same region. This  supports the idea that  the Higgs boson is light, somewhere near 100 GeV, and we missed it at LEP. How can this be?

The basic idea is very simple: if the SM-like Higgs boson decays in a different way than the Higgs boson in the standard model then the usual experimental limits do not apply. Such a Higgs can be light, as predicted from SUSY, gives better agreement with precision electroweak data, and  can even explain the largest excess of Higgs like events at LEP~\cite{Dermisek:2005ar,Dermisek:2005gg,Dermisek:2007yt}. 

In theories beyond the SM the Higgs sector is usually  more complicated and there are typically many other Higgses in addition to the SM-like Higgs boson. For example, there are five Higgs bosons in the MSSM:  light and heavy CP even Higgses, $h$ and $H$, the CP odd Higgs, $A$, and a pair of charged Higgs bosons, $H^\pm$; seven in the next-to-minimal supersymmetric model (NMSSM),  which contains an additional singlet field: three CP even Higgs bosons, $h_{1,2,3}$, two CP odd Higgs bosons, $a_{1,2}$, and a pair of charged Higgs bosons; and there are many simple models with even more complicated Higgs sectors.
Usually we explore parameter space in which the extra Higgses are somewhat heavy -- the so called decoupling limit. 

The decoupling limit is not  the only possibility.
One of the extra Higgses can be light, for example, the singlet CP odd Higgs in the NMSSM. If it is sufficiently light, the SM-like Higgs boson can (and typically would) dominantly decay into a pair of CP odd Higgses and eventually, depending on the mass of the CP odd Higgs boson, into {\bf four} b quarks, {\bf four} $\tau$ leptons, {\bf four} c quarks, {\bf four} $\mu$ leptons,  {\bf four} electrons, {\bf four} light  quarks, or gluons~\cite{Dermisek:2005ar}. Limits on these 4-body final states are weaker than limits on the SM Higgs boson (decaying into $b \bar b$) and currently $4 \tau$ leptons, c quarks,  and light  quarks or gluons  final states allow the SM-like Higgs boson at $\sim 100$ GeV or even lighter~\cite{Chang:2008cw,Dermisek:2010mg}, as is predicted from the best theoretically motivated region of the parameter space in supersymmetric theories, and it also gives much better agreement with precision electroweak data. In addition, the subleading decay mode of the Higgs boson, $h \to b \bar b$, with branching ratio of $\sim 10 \%$ can completely explain the 
largest excess ($2.3 \sigma$) of Higgs-like events at LEP in the
$b\bar b$ final state (for a reconstructed mass $M_{b\bar b}\sim 98$ GeV)~\cite{Dermisek:2005gg}.

Another, perhaps even more interesting 
variation of the above NMSSM scenario is the scenario with a {\bf doublet-like} CP odd Higgs bellow the $b \bar b$ threshold.
For small  $\tan \beta$, $\tan \beta \lesssim 2.5$, this scenario is  the least constrained (and only  marginally ruled out) in the MSSM, and thus easily viable in simple extensions of the MSSM~\cite{Dermisek:2008id,Dermisek:2008sd,Dermisek:2008uu,Bae:2010cd}. Surprisingly, the  prediction from this region is that
all the Higgses resulting from two Higgs doublets: $h$, $H$, $A$ and $H^\pm$ could have been produced already at LEP or the Tevatron, but would have escaped detection because they decay in modes that have not been searched for or the experiments are not sensitive to.
The heavy CP even and the CP odd Higgses could have been produced at LEP in $e^+ e^- \to H A$ but they would avoid detection because the dominant decay mode of $H$,  $H \to ZA$,  has not been searched for. 
The charged Higgs is also very little constrained although it could have been pair produced at LEP or appeared in decays of top quarks produced at the Tevatron. 
The dominant decay mode of the charged Higgs in this scenario is $H^\pm \to W^{\pm} A$ with $A \to \tau^+ \tau^-$ or $A \to c \bar c$.
 In addition, the charged Higgs with  properties emerging in this scenario and the mass close to the mass of the $W$ boson
could explain the $2.8 \sigma$ deviation from lepton universality in $W$ decays measured at LEP~\cite{:2004qh} as pointed out  in Ref.~\cite{Dermisek:2008dq}. 

The main idea is simple, and it can be realized in a variety of other models. For example, in specific little Higgs models the SM-like Higgs boson can dominantly decay into four c quarks~\cite{Bellazzini:2009kw} or four gluons~\cite{Bellazzini:2009xt}. Four body final states of the Higgs boson can also occur in composite Higgs models~\cite{Gripaios:2009pe}, and models for dark matter~\cite{Mardon:2009gw} among others.  For a review of other scenarios and references, see Ref.~\cite{Chang:2008cw}.

The situation can be even more complicated if there are several lighter Higgses. In this case the SM-like Higgs boson can cascade decay into the lightest one, and eventually, depending on the mass of the lightest Higgs boson,  the SM-like Higgs would decay into a large number of light jets, or a large number of muons, or a large number of electrons, or large number of soft photons~\cite{Dermisek:2006wr,superB-talk,Dermisek:2009si}. Such events would be quite spectacular. Some of these signatures were recently studied in Ref.~\cite{Falkowski:2010cm}. Additional level of complexity arises when more Higgs boson share the coupling to the Z boson and there is not a single SM-like Higgs boson, see {\it e.g.}~\cite{Dermisek:2007ah}.

The simplest models allowing the SM-like Higgs boson at $\sim 100$ GeV are still those with  4-body decay modes. The NMSSM scenario with $h \to 4 \tau$  has been recently studied  and searched for at a variety of experiments. In the rest of this talk we will focus on this scenario.

\section{Experimental searches and constraints}

The first constraints on these scenarios came from CLEO~\cite{:2008hs}. 
The light CP odd Higgs can be produced  in Upsilon decays, $\Upsilon \to A \gamma$, and the predicted branching ratio from the NMSSM model typically varies between $few \times 10^{-4}$ for the CP odd Higgs mass close to $2 \tau$ threshold, and $10^{-7}$ for the CP odd Higgs mass close to the Upsilon mass~\cite{Dermisek:2006py}.
At present, the strongest constraints come from BaBar~\cite{:2009cp, Aubert:2009cka} that sets limits  $B (\Upsilon \to A \gamma) \lesssim 10^{-5}$ (slightly varying around this value for different masses of the CP odd Higgs boson).
In order to satisfy these limits it is typically required that  $m_A \gtrsim 8$ GeV for $\tan \beta > 3$. For smaller $\tan \beta$ these limits become weaker  as $A \to c \bar c$ channel  becomes more  important. More details about the impact of these searches on the NMSSM parameter space can be found in Ref.~\cite{Dermisek:2010mg}.

Searches for $h \to aa$ were performed and are in progress at the Tevatron~\cite{Abazov:2009yi,Wilbur}. These searches are not sensitive yet to the SM-like Higgs boson at 100 GeV. Nevertheless it is interesting to note that it is advantageous to search for a subleading decay mode of one of the CP odd Higgs bosons, $a \to \mu^+ \mu^-$, as suggested in Ref.~\cite{Lisanti:2009uy}. The rate for $h \to aa \to 2 \tau 2\mu$ is suppressed by a factor of $\sim m_\mu^2/m_\tau^2$ compared to $h \to aa \to 4 \tau$,  but these events are much cleaner and one can reconstruct the mass of the CP odd Higgs boson. This search mode is very promising at the LHC where one expects about 500 events in 1 fb$^{-1}$ of data (for 14 TeV center of mass energy)~\cite{Lisanti:2009uy}.

The most important constraints on this scenario come from the recent search at Aleph~\cite{Schael:2010aw}. Naively this search rules out the SM-like Higgs that decays into four $\tau$ leptons up to 107 GeV. Note however, that this search places limits on 
$$\xi^2 = \frac{\sigma(e^+ e^- \to Z h)}{\sigma (e^+ e^- \to Z h_{SM})} \times B(h \to aa) \times B(a \to \tau^+ \tau^-)^2,$$
and thus it is very sensitive to $B(a \to \tau^+ \tau^-)$, which in the NMSSM is never equal (or even close) to 100\%. For large $\tan \beta$, depending on the mass of the CP odd Higgs boson, $B(a \to \tau^+ \tau^-)$ is between 0.9 and 0.7 because of the sizable branching ratio of $a\to gg$. In addition, for smaller $\tan \beta$ a new decay mode, $a \to c \bar c$, becomes important.  
Folding in realistic branching ratios of $a \to \tau^+ \tau^-$ one finds that Aleph limits allow the SM-like Higgs boson at 100 GeV (or slightly lighter) for any $\tan \beta > 3$ only when $m_a \gtrsim 9.5 $ GeV, and generically for $\tan \beta <2$~\cite{Dermisek:2010mg}.

\section{Prospects at the LHC}

Recently we have been working on new ways to search for these scenarios. One question was whether we can directly observe the light CP odd Higgs (without relying on producing heavier Higgses that decay into the CP odd Higgs). The direct production cross section of the light CP odd Higgs boson in gluon fusion channel  is, depending on the mass of the CP odd Higgs boson and $\tan \beta$, between  1 and 100 nb   at the Tevatron, and between 10 and 1000 nb at  the LHC~\cite{Dermisek:2009fd}. In spite of the large production cross section, picking up the signal on huge background is a serious problem. 
If one looks for the dominant decay mode, $a \to  \tau^+ \tau^-$, then it is basically hopeless, however searching for the subleading decay mode, $a \to \mu^+ \mu^-$,  is actually very promising. We still have to deal with huge background from Drell-Yan production of $\mu$-pairs, and semileptonic b and c decays. Nevertheless, based on existing analyses we showed that current data sets would allow CDF and D0 to improve on limits from Babar, especially for masses of the CP odd Higgs boson close to the mass of the Upsilon~\cite{Dermisek:2009fd}. 
At the LHC this is a very promising search mode and already the first  fb$^{-1}$ of data might provide an evidence for, or the discovery of  the light CP odd Higgs.
More details, and predictions for  the integrated luminosity needed for the discovery as a function of the mass and couplings of the light CP odd Higgs boson can be found in Ref.~\cite{Dermisek:2009fd}.

Looking for subleading decay modes of the light CP odd Higgs might be also advantageous in searches for the charged Higgs. 
If the CP odd Higgs boson has a significant dublet component than the charged Higgs is generically  light, typically lighter  than the top quark.
Depending on the  mass of the charged Higgs and $\tan \beta$ the  $B(t \to H^+ b)$  can go up to 40\% for $\tan \beta =1$ and $m_{H^\pm} = 80$ GeV dropping very fast with increasing  $\tan \beta$ and increasing the mass of the charged Higgs. 
 The dominant decay mode of the charged  Higgs  is $H^+ \to W^+ A$ and it was not previously searched for (till recently). The CDF recently performed the search and set the upper limit on $B( t \to H^+ b) \times B(H^+ \to W^+ A) \times B( A\to \tau^+ \tau^-)$ to about 10\%~\cite{CDF_charged_higgs} constraining only a small region of the parameter space. 
 
 The search for the charged Higgs decaying into $W^\pm A$ will be relatively easy  at the LHC  which is the top quark factory. We might again look for a subleading decay mode $a \to \mu^+ \mu^-$ that we  cannot afford  at the Tevatron because the rate would be too small. In addition, we can look for events in which one of the $W$ bosons decays into $\mu \nu$, resulting in 3-muon events with properties that easily stand out from the background.   We expect about 30 events of this type with 1 fb$^{-1}$ of data at the LHC~\cite{DLR}.

\vspace{.3cm}

In conclusion, four body decay modes: $4 \tau$, $4c$, $4q$, and $4g$ are the simplest possibilities allowing the SM-like Higgs boson at $\sim 100$ GeV. For the  $h\to aa \to 4\tau$ scenario in the NMSSM, searching for  dominant decay modes typically requires many tens of fb$^{-1}$ of data. However, searching for the subleading decays modes is very promising with early data at the LHC. Especially searches for
\begin{itemize}
\item $gg \to h \to aa \to 2\tau 2\mu$~\cite{Lisanti:2009uy},
\item $gg \to a \to 2\mu$~\cite{Dermisek:2009fd},
\item $t \to H^+ b, \;\;\;\; H^+ \to W^+ a, \;\;\;\; a \to \mu^+ \mu^- $~\cite{DLR}
\end{itemize}
 could lead to an evidence or discovery already with 1 fb$^{-1}$ of data.

\section*{Acknowledgments}
I would like to thank J. Gunion, H.D. Kim, E. Lunghi and A. Raval for collaboration on projects this talk is based on.


\begin{footnotesize}

\end{footnotesize}


\end{document}